\documentclass[twoside]{dis04}

\def\runauthor{A. Bruni, M. Diehl, F.-P. Schilling}
\def\shorttitle{Summary of WG B: Diffraction and Vector Mesons}

\usepackage{epsfig}

\newcommand{\pom}{{I\!\!P}}

\newcommand{\xpom}{x_\pom}

\newcommand{\gapprox}{\stackrel{>}{_{\sim}}}
\newcommand{\alphapom}{\alpha_{_{\rm I\!P}}}

\newcommand{\sigrd}{\sigma_r^{D(3)}}

\newcommand{\gsim}{\raisebox{-4pt}{$
\,\stackrel{\textstyle >}{\sim}\,$}}

\newcommand{\bc}{\begin{center}}
\newcommand{\ec}{\end{center}}
\newcommand{\bi}{\begin{itemize}}
\newcommand{\ei}{\end{itemize}}
\newcommand{\be}{\begin{enumerate}}
\newcommand{\ee}{\end{enumerate}}

\def\Journal#1#2#3#4{{#1} {\bf #2} (#3) #4}

\def\NPB{{\em Nucl. Phys.}   {\bf B}}
\def\PLB{{\em Phys. Lett.}   {\bf B}}
\def\PRL{\em Phys. Rev. Lett.}
\def\PRD{{\em Phys. Rev.}    {\bf D}}
\def\ZPC{{\em Z. Phys.}      {\bf C}}
\def\EJC{{\em Eur. Phys. J.} {\bf C}}

\newcommand{\etal}{{\em et al.}}

\begin{document}

\title{
SUMMARY OF WORKING GROUP B: \\
DIFFRACTION AND VECTOR MESONS
}

\author{ALESSIA BRUNI$^1$, MARKUS DIEHL$^2$, FRANK-PETER SCHILLING$^2$}

\address{
1. INFN Bologna, I-40156 Bologna, Italy \\
2. Deutsches Elektronen-Synchroton DESY, D-22603 Hamburg, Germany\\
E-mails: alessia@mail.desy.de, mdiehl@mail.desy.de, fpschill@mail.desy.de}

\maketitle


\abstracts{ 
  We summarise the contributions presented in working group
  B: ``Diffraction and Vector Mesons''.
}


\section{Introduction} 

The understanding of diffractive lepton-hadron or hadron-hadron
interactions at high energies, where at least one of the beam hadrons
stays intact and loses only a small fraction of its incident momentum,
still represents one of the main challenges in Quantum
Chromodynamics. Diffraction is being extensively studied both at HERA
and the TEVATRON, and there is a growing community planning to
continue this research at the LHC.

The current experimental data as well as future plans have been
reviewed during the sessions in Working Group B,
where in
total 21 experimental talks were presented.  There were 19
theoretical talks, 11 of which dealt with the topic of saturation.
In the following, contributions will be summarised which cover
diffraction at HERA (section \ref{sec:hera}) and the TEVATRON (section
\ref{sec:tevatron}), vector meson production and DVCS (section
\ref{sec:vmdvcs}), the phenomenology of saturation (section
\ref{sec:sat}), and future experimental opportunities (section
\ref{sec:lhc}).


\section{Diffraction at HERA}
\label{sec:hera}

\subsection{Inclusive Diffraction}
\label{sec:herainc}

\begin{figure}
\centering
\epsfig{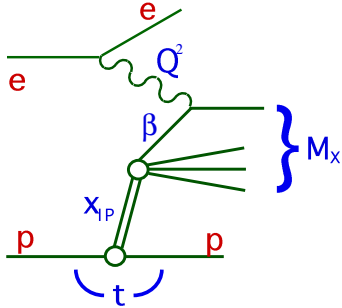}
\caption{\label{fig:feyn} Illustration of a diffractive DIS event.}
\end{figure}

In diffractive deep-inelastic scattering at HERA
(Figure~\ref{fig:feyn}) the virtual photon $\gamma^*$ emitted from the
beam electron provides a point-like probe to study the structure of
the diffractive exchange, similarly to ordinary DIS probing proton
structure.  Experimentally, diffractive events are identified either
by tagging the elastically scattered proton in {\em Roman pot}
spectrometers $60-100\rm\ m$ downstream from the interaction point or
by properties of the hadronic final state, for example by a large
rapidity gap without particle production between a central hadronic
system and the proton beam direction.  The diffractive {\em reduced
cross section} $\sigma_r^{D(4)}$ is defined as
\begin{eqnarray}
\frac{d^4\sigma^{ep\rightarrow eXp}}{dx_\pom \ dt \ d\beta \ dQ^2}=
\frac{2\pi\alpha^2 Y_+}{\beta Q^4}
\sigma_r^{D(4)}(x_\pom,t,\beta,Q^2) \ .
\end{eqnarray}
Here $\xpom$ is the longitudinal momentum loss of the incident proton,
$t$ is the squared four-momentum transfer at the proton vertex,
$\beta$ is the momentum fraction of the quark struck by the photon
with respect to the diffractive exchange (i.e.\ the equivalent of $x$
in ordinary DIS), and $Q^2$ is the photon virtuality.  One further has
$Y_+=1+(1-y)^2$ in terms of the usual inelasticity variable $y$.  The
reduced cross section $\sigma_r^{D(4)}$ is related to the diffractive
structure functions $F_2^D$ and $F_L^D$ by $\sigma_r^D =
F_2^D-(y^2/Y_+) F_L^D$.  Except at the highest values of $y$, one has
$\sigma_r^D=F_2^D$ to a very good approximation. If the outgoing
proton is not detected, the measurements are integrated over $t$,
i.e.\ $\sigma_r^{D(3)}(x_\pom,\beta,Q^2) = \int {\rm d}t \
\sigma_r^{D(4)}(x_\pom,t,\beta,Q^2)$.

Both H1 and ZEUS \cite{kapishin,ruspa} have presented recent precise
measurements of $\sigrd$.  The H1 measurement selects diffractive
events by requiring a large rapidity gap and covers low, medium and
high $Q^2$, with $1.5<Q^2<1600 \rm\ GeV^2$ and $\xpom<0.05$
\cite{h1f2d3}.  The data are consistent with measurements using the H1
Forward Proton Spectrometer FPS to tag the diffractively scattered
proton directly \cite{h1fps}.  ZEUS selected diffractive events by
using both the so-called ``$M_X$ method'' ($2.2<Q^2<80 \rm\ GeV^2$)
and the Leading Proton Spectrometer LPS ($0.03<Q^2<100 \rm\ GeV^2$)
\cite{zeusf2d}.  The ZEUS LPS data cover a large range in $\xpom$ up
to 0.1, a region relevant for the comparison with the Tevatron data
(see section~\ref{sec:tevatron}).

\begin{figure}
\centering
\epsfig{file=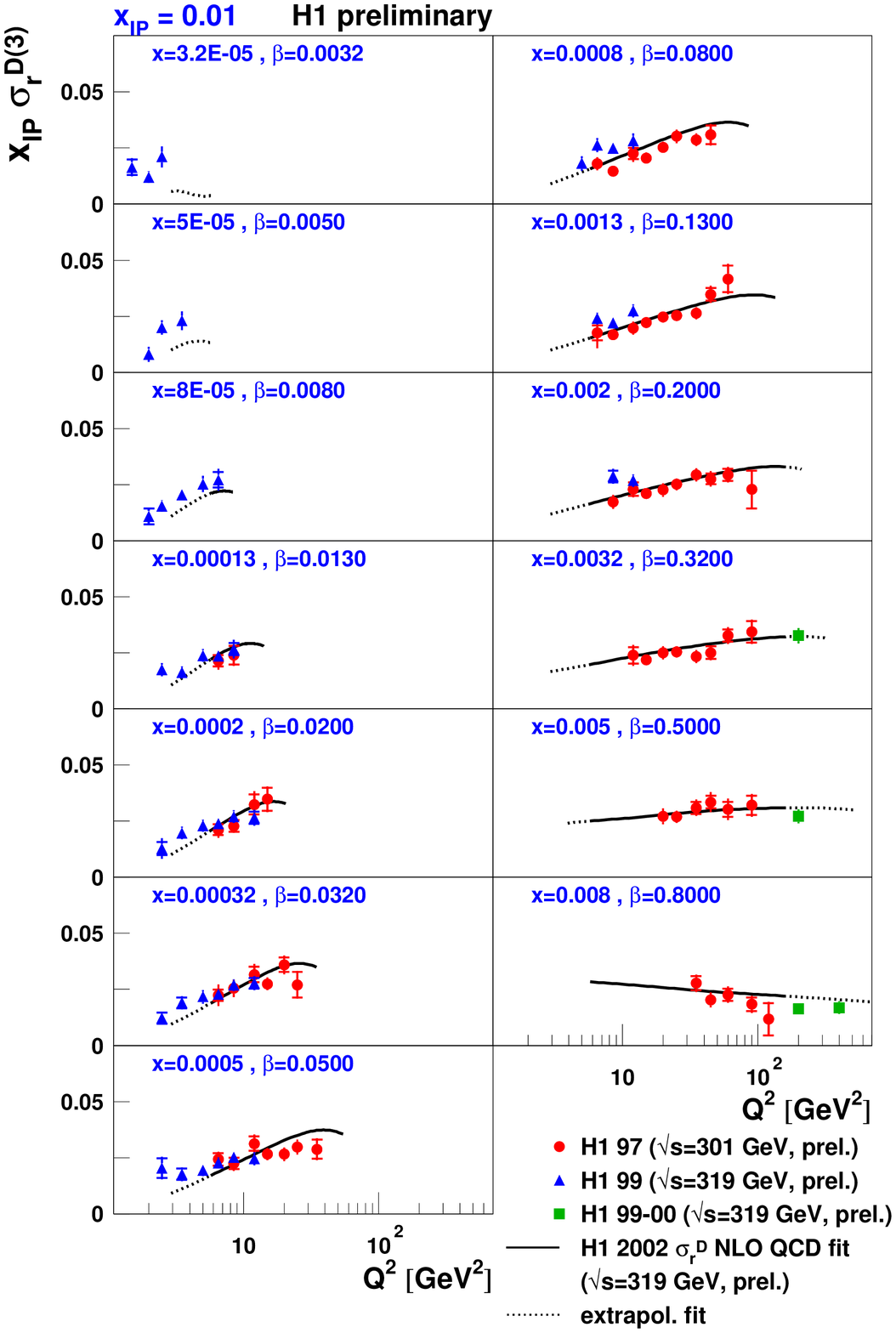,width=0.85\linewidth}
\caption{\label{fig:h1q2dep} H1 measurements of the $Q^2$ dependence
of $x_\pom\, \sigrd$.  }
\end{figure}

\begin{figure}
\centering
\epsfig{file=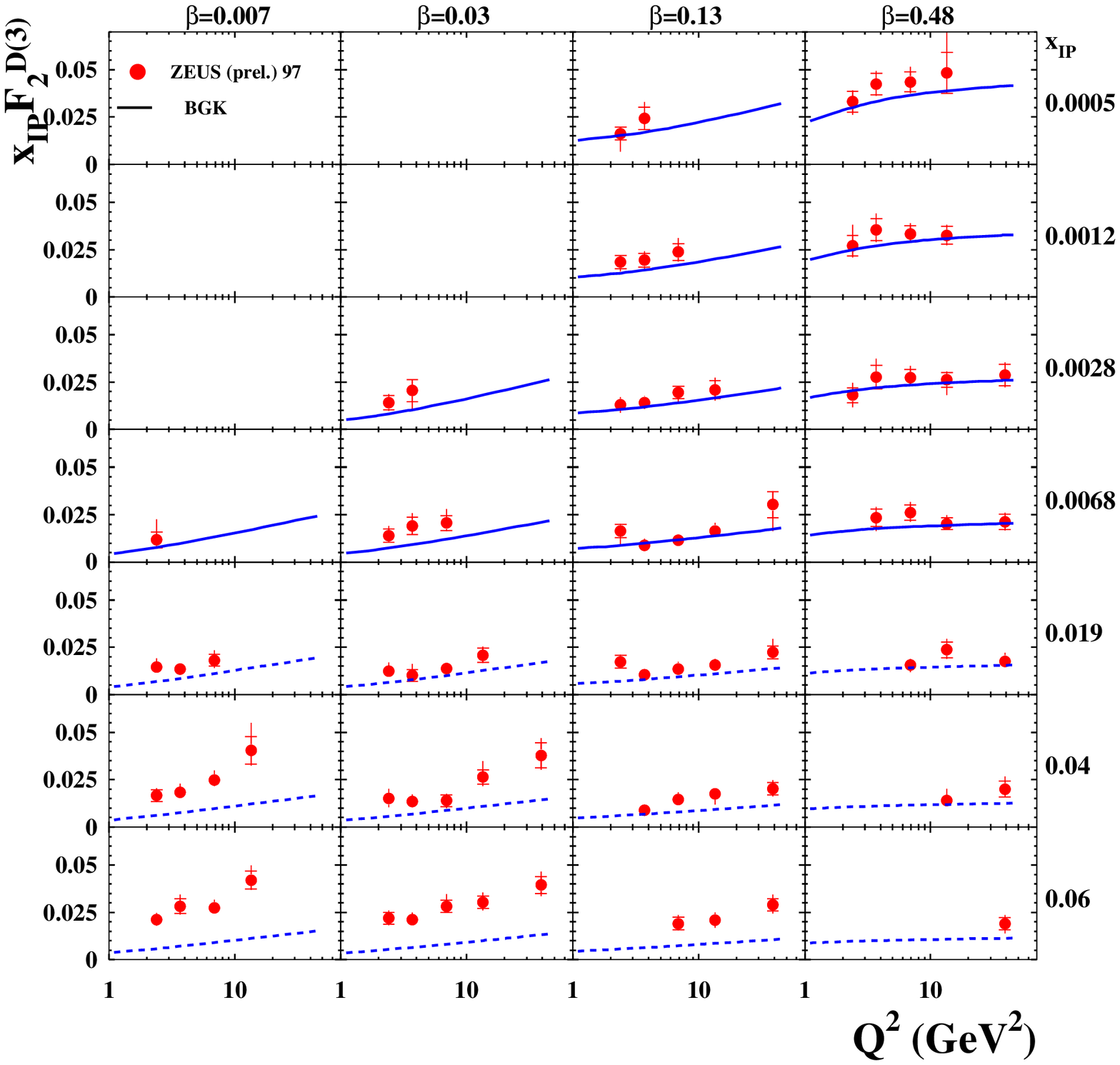,width=0.95\linewidth} \\
\caption{\label{fig:zeusq2dep} ZEUS measurements of the $Q^2$
dependence of $x_\pom\, F_2^{D(3)}$.  }
\end{figure}

The H1 and ZEUS data exhibit clear positive $Q^2$ scaling violations
indicative of a large gluonic contribution to the exchange (see
Figures~\ref{fig:h1q2dep} and \ref{fig:zeusq2dep}). The ratio of the
diffractive to the inclusive DIS cross section is observed to be flat
except at the highest $\beta$.  For $Q^2\gapprox10\rm\ GeV^2$, the
$\xpom$ dependence of the cross section can be parameterized in terms
of an effective Pomeron intercept $\alphapom(0)\sim 1.2$, which is
significantly larger than the soft Pomeron value of $1.08$. Resolving
the issue of a possible variation of $\alphapom(0)$ with $Q^2$ in
inclusive diffractive DIS needs further experimental input.

\begin{figure}
\centering
\epsfig{file=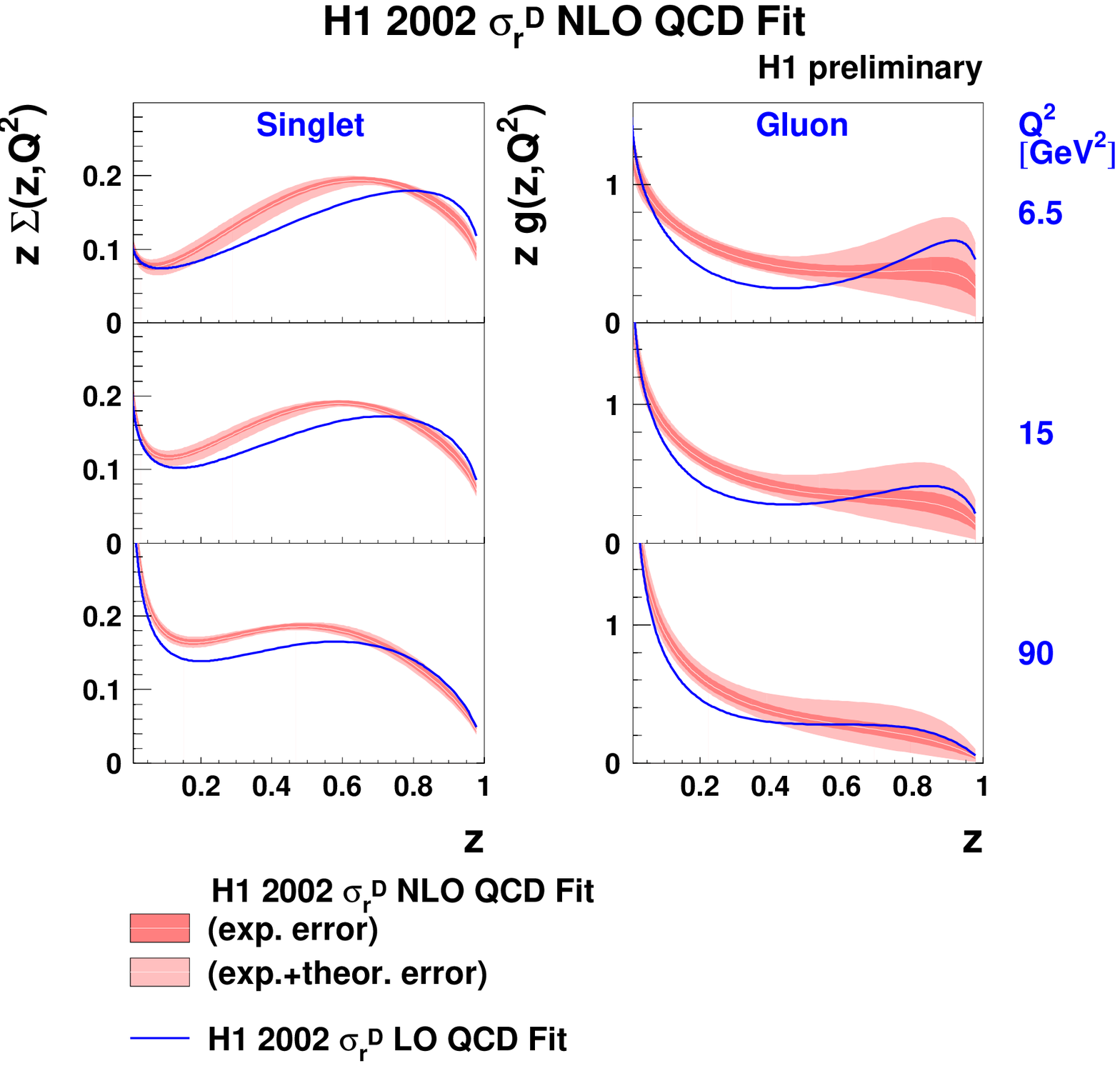,width=0.75\linewidth}
\caption{\label{fig:h1pdfs} Diffractive parton distributions obtained
from the H1 NLO QCD fit.}
\end{figure}

\begin{figure}
\centering
\epsfig{file=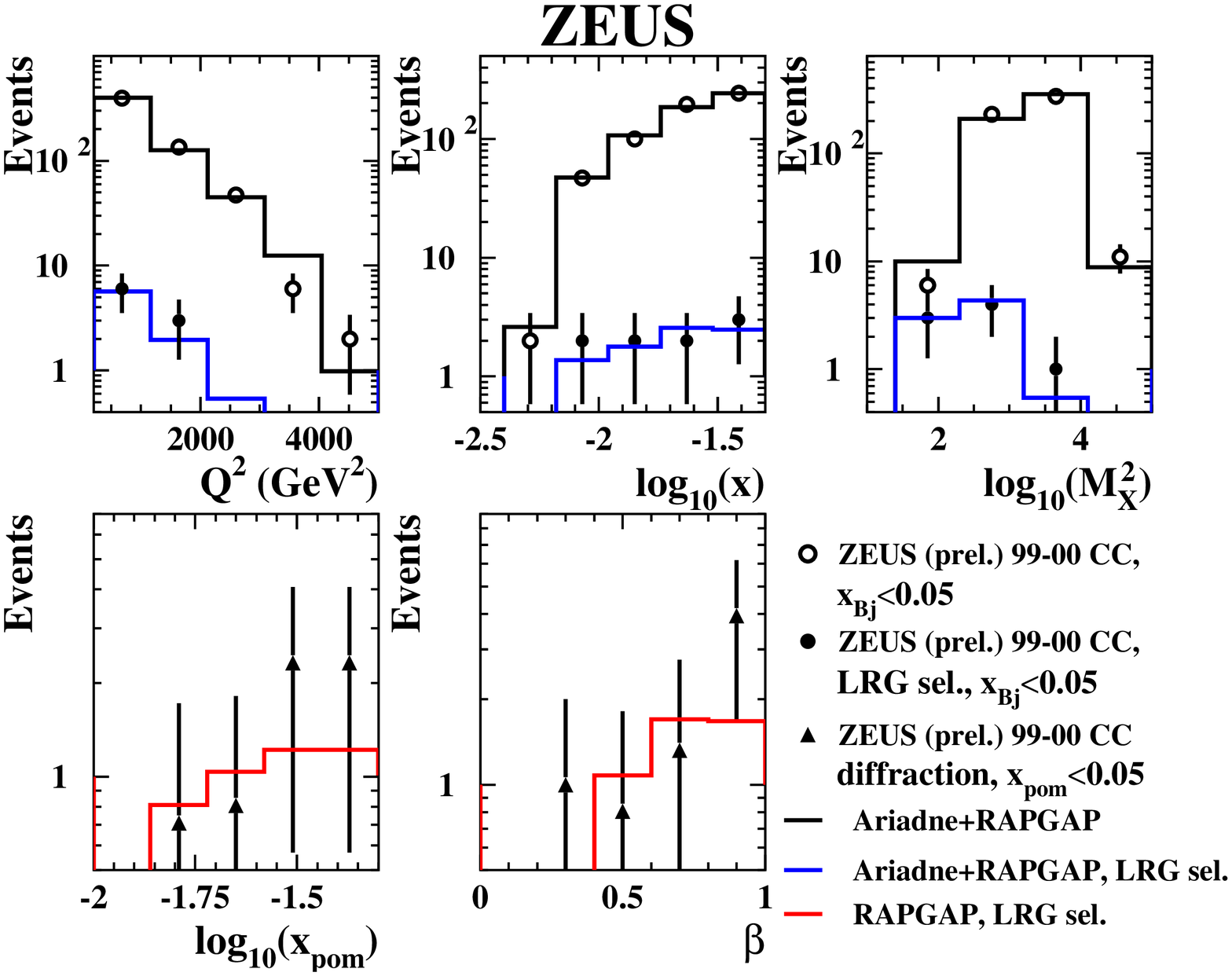,clip=,width=0.82\linewidth}
\caption{\label{fig:zeuscc} Distributions for charged current events
with and without a large rapidity gap as observed by ZEUS.}
\end{figure}

Applying the QCD factorization theorem for diffractive DIS
\cite{collins}:
\begin{equation}
\frac{{\rm d^2} \sigma(x,Q^2,x_\pom,t)^{\gamma^*p\rightarrow pX}}
{{\rm d} x_\pom \ {\rm d} t} \ = \
\sum_i \int_x^{x_\pom}{\rm d}z \
\hat{\sigma}^{\gamma^*i}(x,Q^2,z) \
p_i^D(z,Q^2,x_\pom,t) \ ,
\label{equ:diffpdf}
\end{equation}
both collaborations have performed next-to-leading order (NLO) QCD
fits to their diffractive DIS data \cite{kapishin,ruspa}. The NLO
diffractive parton distributions $p_i^D$ extracted from the H1 data
are shown in Figure \ref{fig:h1pdfs} (they will be referred to as H1
2002 fit in the following). The shape of the diffractive PDFs
extracted by ZEUS is less well constrained (only the statistically
limited LPS data with $x_\pom < 0.01$ were used), but using in
addition diffractive charm data \cite{Chekanov:2003gt} further
constrains the diffractive gluon distribution.  In the fits of both H1
and ZEUS, the momentum fraction of the diffractive exchange carried by
gluons is determined to be about $80\%$ to $85\%$.  The diffractive
PDFs can be used to test QCD factorization in diffraction by
predicting jet and charm cross sections (sections \ref{sec:herafs} and
\ref{sec:tevatron}).

At the workshop, ZEUS has reported the observation of events with a
large rapidity gap in charged current interactions at high $Q^2$
\cite{klimek}.  In $61 \rm\ pb^{-1}$ of data, 9 events are observed
for $Q^2>200 \rm\ GeV^2$ and $\xpom<0.05$, corresponding to a cross
section of $\sigma^{\rm cc-D}=0.49 \pm 0.20 {\rm(stat.)}\pm 0.13 {\rm
(syst.)} \rm\ pb$, see Figure \ref{fig:zeuscc}.  With the Monte Carlo
RAPGAP and the H1 2002 diffractive PDFs, one expects $5.6$ events over
a background of $2.1$ events.
 
ZEUS has also presented \cite{sacchi} a measurement of leading neutron
production for $Q^2\sim 0$ in the kinematic range $0.05<|t|<0.425 \rm\
GeV^2$ and $0.6<x_L<0.925$, where $x_L=E_n/E_p$
\cite{zeuspion}. Within the reggeised one-pion exchange model, the
data have been used to extract the slope of the pion trajectory as
$\alpha'_\pi = 1.39 \pm 0.16 \pm 0.26 \rm\ GeV^{-2}$, supporting the
applicability of this model in the reaction $\gamma p \rightarrow
nX$.  Results were also presented \cite{sacchi} on measurements of DIS
with a leading proton \cite{zeuslp}.

\subsection{Diffractive Final States}
\label{sec:herafs}


Since QCD hard scattering factorization holds for diffractive DIS,
calculations based on diffractive PDFs extracted from inclusive
measurements of $\sigrd$ can predict cross sections for diffractive
final states such as dijets or charm production.  However, the
factorization proof of Collins \cite{collins} does not hold in the
case of photoproduction, where the photon is quasi-real ($Q^2\sim 0$)
or in the case of diffractive hadron-hadron scattering.

In fact, it has been known for a few years that diffractive PDFs
extracted from HERA data overestimate the rate of diffractive dijet
events at the TEVATRON by one order of magnitude \cite{cdfjets}. This
breakdown of factorization can be attributed to additional soft
interactions between the beam hadrons and their fragments.  Such
interactions explicitly appear in the analysis of the factorization
theorems: they cancel out in $\gamma^* p$ processes but not in
hadron-hadron collisions \cite{collins}.  They are expected to reduce
the rate of observed diffractive events ({\it rapidity gap survival
probability}) and have been modeled in soft physics approaches.  It is
thus interesting to study at HERA the rate of diffractive dijet
photoproduction events, since the real photon can act similarly to the
second hadron in $p\bar{p}$ collisions.

\begin{figure}
\epsfig{file=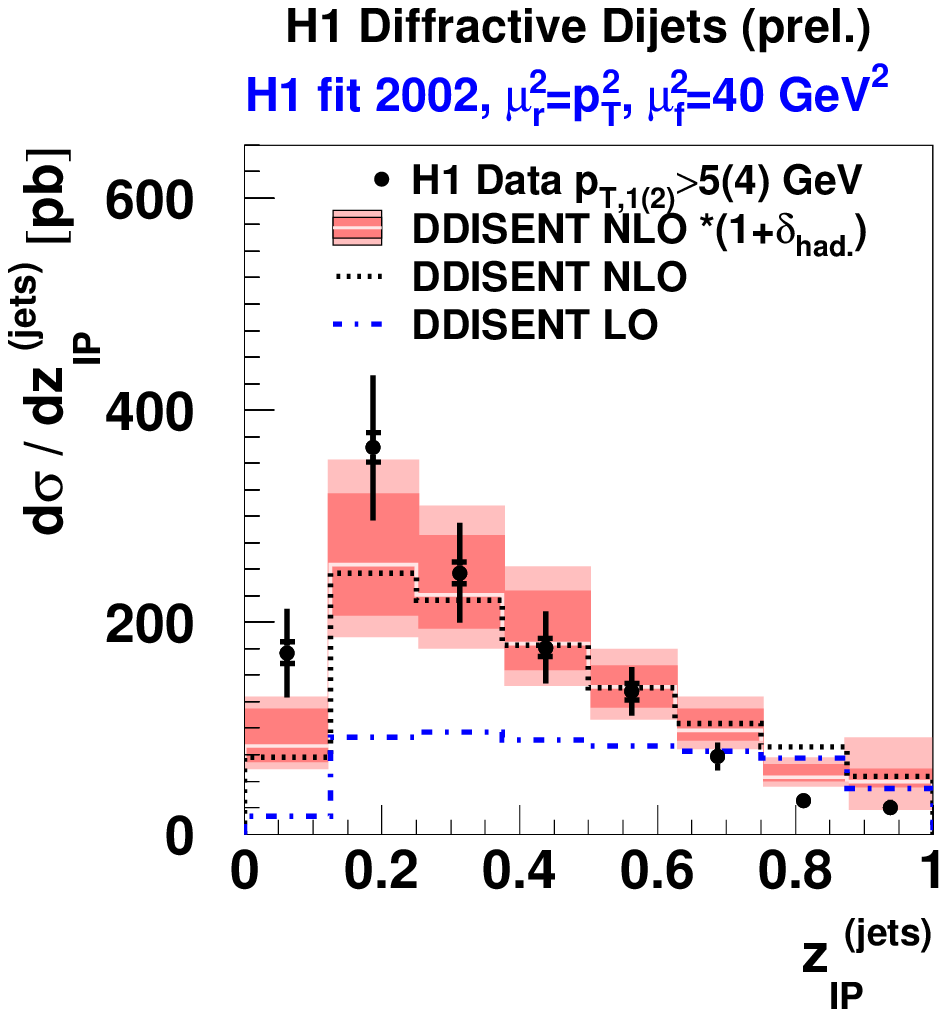,clip=,width=0.47\linewidth}
\hspace{0.2cm}
\epsfig{file=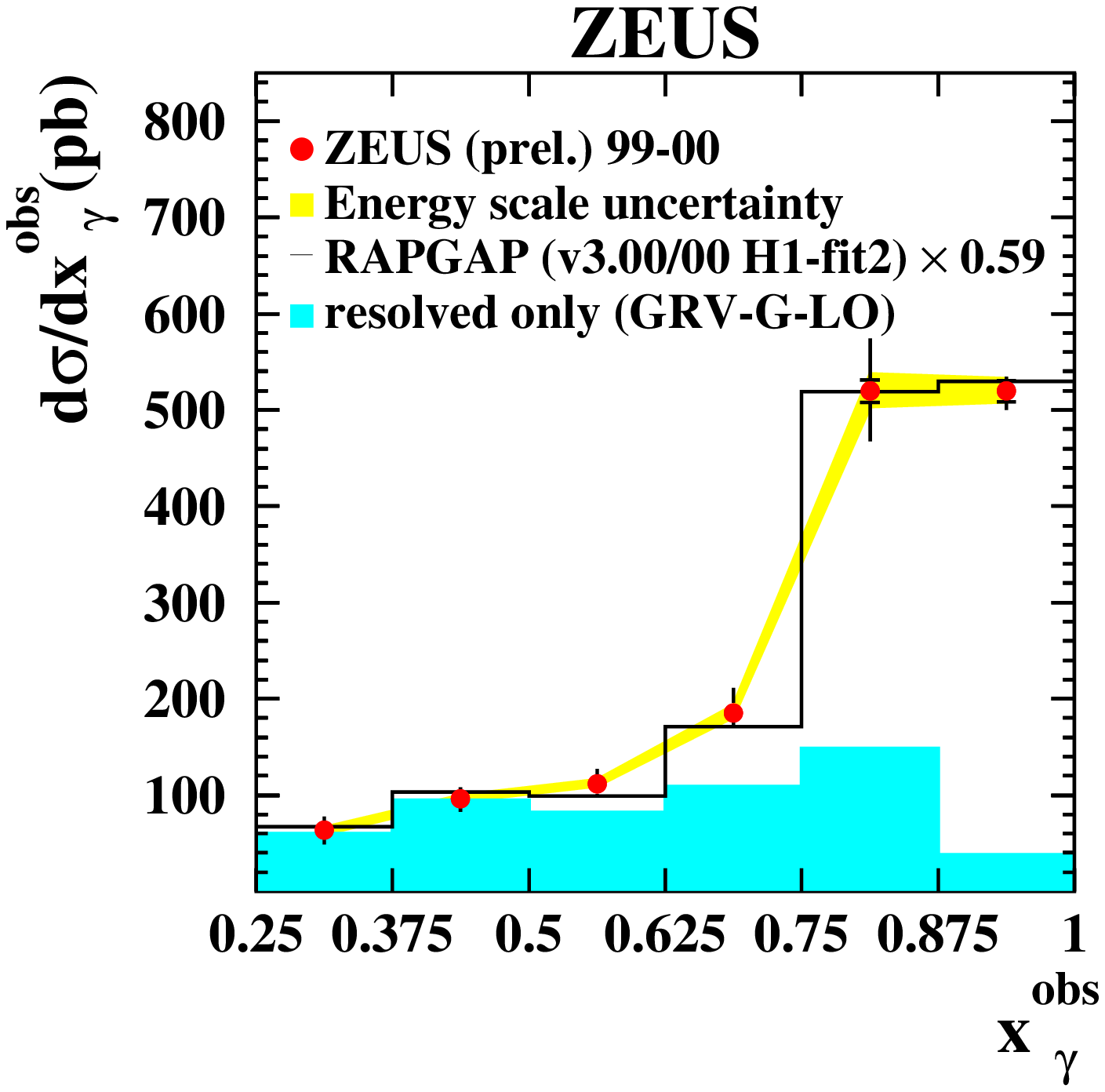,clip=,width=0.47\linewidth} \\
(a) \hspace{0.45\linewidth} (b)
\caption{\label{fig:jets} (a) H1 diffractive DIS dijet cross section,
compared with NLO calculations.  (b) ZEUS diffractive dijet cross
section in photoproduction, compared with LO calculations.  }
\end{figure}

At the workshop, H1 has presented \cite{schaetzel} new results on
comparisons between dijet and charm production in diffractive DIS with
predictions based on the H1 2002 diffractive PDFs (Figure
\ref{fig:h1pdfs}).  An important development is that these comparisons
are now done by calculating the hard cross sections to NLO accuracy,
i.e.\ to the same precision as in standard DIS.  The diffractive dijet
cross section for $4<Q^2<80 \rm\ GeV^2$, $x_\pom<0.05$ and
$p_{T,jet}^{1(2)}>5(4) \rm\ GeV$ is shown in Figure~\ref{fig:jets}(a),
compared with NLO calculations based on the H1 NLO diffractive PDFs.
Good agreement is observed, in support of QCD factorization.  Similar
conclusions can be drawn from the study of diffractive charm
production in DIS, presented by both H1 \cite{schaetzel,h1nlo} and
ZEUS \cite{vlasov,zeusdstar} .

In the case of diffractive dijet photoproduction, both HERA
collaborations have presented very interesting results: ZEUS has shown
\cite{kagawa} new cross section measurements in the kinematic range
$Q^2<1 \rm\ GeV^2$, $\xpom<0.035$ and $p_{T,jet}^{1(2)}>7.5(6.5) \rm\
GeV$ (Figure \ref{fig:jets}(b)). The measurements are compared with
the leading order (LO) Monte Carlo program RAPGAP using the previous
H1 fits 2 and 3 of diffractive PDFs \cite{h1f2d97}. Good agreement in
both shape and normalization is achieved if the prediction is scaled
by a factor $0.6$. In contrast, H1 \cite{schaetzel,h1gpjets} finds
good agreement at LO without having to rescale the prediction when the
more recent H1 2002 diffractive PDFs \cite{h1f2d3} are used. The two
results are not contradicting each other, given that the diffractive
gluon distribution in the H1 2002 fit is smaller than in the previous
H1 fits 2 or 3 by a factor similar to the rescaling factor 0.6 needed
in the ZEUS analysis.  When comparing photoproduction dijet data with
LO calculations using the 2002 diffractive PDFs from H1, both the ZEUS
and H1 results are thus consistent with a rapidity gap survival
probability close to 1, in contrast to the findings at the TEVATRON.

Kramer \cite{kramer} has presented an analysis of diffractive dijet
photoproduction at NLO, using the diffractive PDFs from the new H1 NLO
analysis as an input.  Compared with the H1 data, the calculation is
found to give too large a cross section.  Agreement with the data is
found when introducing a suppression factor of 0.34 for the
\emph{resolved} photon part in the calculation (whose value was taken
from a model evaluation of the gap survival probability for this
process in \cite{kaidalov}).  After this conference, both H1 and ZEUS
have presented comparisons of their measurements with NLO calculations
and found that a \emph{global} suppression factor is required to
describe the data, i.e.\ a suppression of both the resolved and the
direct photon part \cite{h1ichep,zeusichep}.  The origin of the
discrepancy between these different studies remains to be clarified,
as well as their relation with the comparisons at LO.

In a further contribution to the workshop, results from ZEUS
\cite{zeuslcwf} on the process $\gamma \gamma \rightarrow \mu\mu$ were
presented \cite{ukleja}, which are sensitive to the electromagnetic
light-cone wave function of the photon.


\section{Diffraction at the TEVATRON}
\label{sec:tevatron}

Results on diffraction at the TEVATRON were presented at the workshop
in two contributions by the CDF \cite{terashi} and D0 \cite{edwards}
collaborations.

\begin{figure}
\centering
\epsfig{file=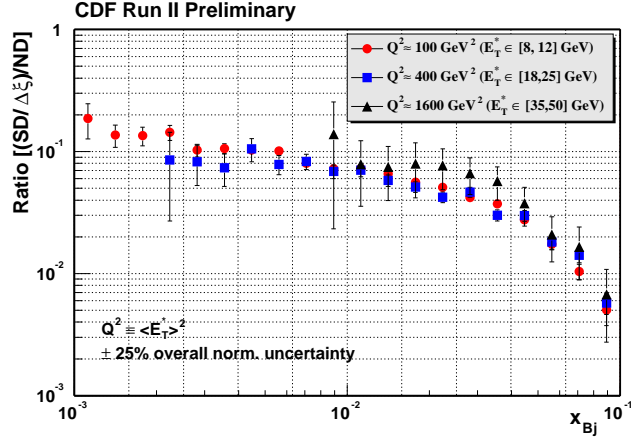,clip=,width=0.65\linewidth} \\[2em]
\caption{\label{fig:cdfa} Ratio of single- to non-diffractive dijet
production from CDF.}
\end{figure}

\begin{figure}
\centering
\epsfig{file=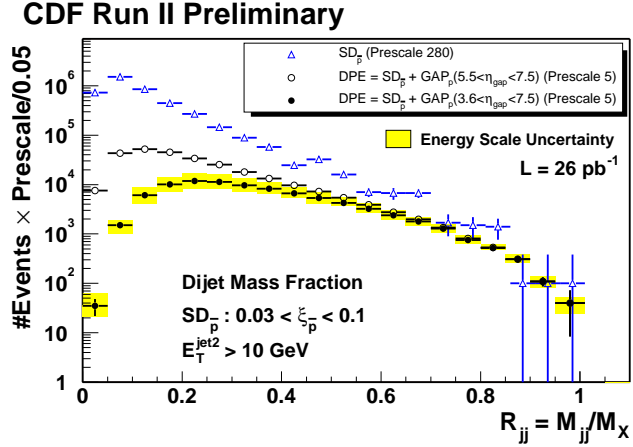,clip=,width=0.65\linewidth}
\caption{\label{fig:cdfb} Fractional invariant mass contained in the
dijet system for single- and double-diffractive events from CDF.  }
\end{figure}

CDF \cite{terashi} has determined the ratio of single-diffractive to
non-diffractive dijet events \cite{cdfjets} for $p_{T,jet}>7 \rm\ GeV$
and $0.035<\xpom<0.095$, which is found to be one order of magnitude
smaller than what is expected using the diffractive PDFs from HERA.
These results are consistent with CDF studies of diffractive $J/\Psi$
production. However, the ratio of double- to single-diffractive dijets
is found to be about a factor $5$ larger than the ratio of single- to
non-diffractive dijets, suggesting that there is at most a small extra
suppression when going from one to two rapidity gaps in the event.

Within the approach of Good and Walker to diffraction, Bialas
\cite{bialas} has shown that in fact there is no suppression when
going from one to two gaps in models where the interactions
responsible for factorisation breaking are restricted to purely
elastic scattering (neglecting e.g.\ interactions involving proton
dissociation, which are included in more refined approaches).

With new detectors (``MiniPlug Calorimeter'') added to the existing
Roman Pot devices, CDF has now improved capabilities for detecting
diffractive events in Run~II. Using the new Run II data, CDF could
reproduce the results obtained in Run~I. In addition, the jet
measurement is now performed in several $p_{T,jet}$ intervals (Figure
\ref{fig:cdfa}). The ratio of single- to non-diffractive dijets shows
no significant $p_T^2$ dependence in the range $100<p_T^2<1600 \rm\
GeV^2$.

Run II data are also being used by CDF to search for exclusive dijet
(Figure~\ref{fig:cdfb}) and $\chi_c$ production in double diffractive
events, with first results presented at this workshop.  The
measurement of the cross sections of these processes is considered to
provide important calibration points necessary to normalize model
predictions for diffractive Higgs production at the LHC (see section
\ref{sec:lhc}).

In contrast to Run I, the D0 detector includes Roman Pot spectrometers
in Run~II \cite{edwards}. The outgoing beam pipes for both $p$ and
$\bar{p}$ are equipped with in total 9 spectrometers composed of 18
Roman Pots. The new detectors have been used for an initial
measurement of the elastic $t$ slope and are expected to provide a
wealth of diffractive data.


\section{Vector Meson Production and DVCS}
\label{sec:vmdvcs}

In a joint session of the diffractive and the spin working group,
vector meson production and deeply virtual Compton scattering (DVCS)
were discussed.  Contributions from HERMES to this session
\cite{ellinghaus,borissov} are summarised in \cite{spinsummary}.

\begin{figure}
\centering
\epsfig{file=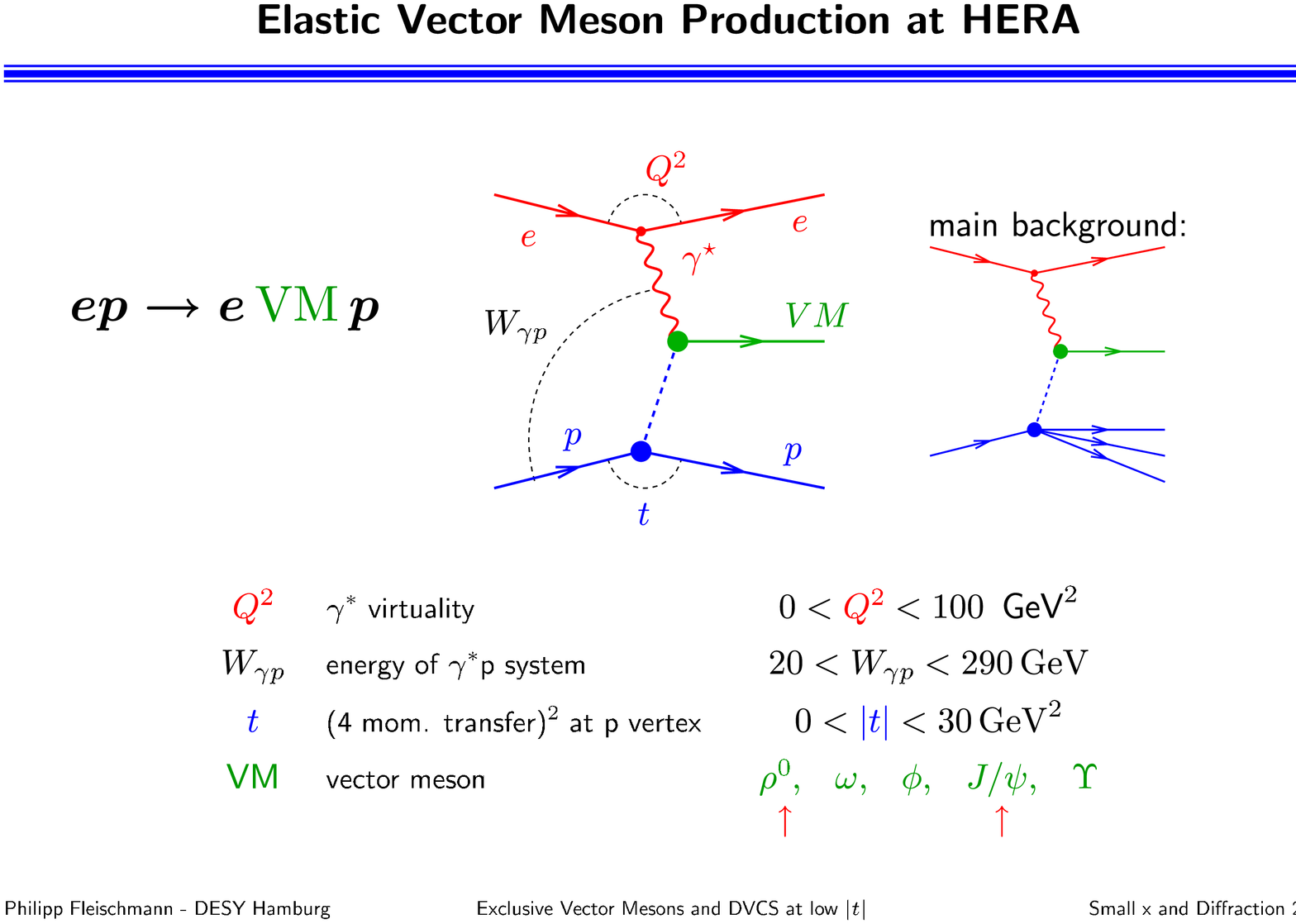,clip=,angle=270,width=0.49\linewidth}
\caption{\label{fig:vm} Diagram of elastic vector meson production at
HERA.  }
\end{figure}

Diffractive vector meson production at HERA, $e+p \rightarrow e+ V+Y$,
where $V$ is the vector meson and $Y$ is either a proton (``elastic'')
or a low-mass proton dissociation system, provides a clean laboratory
to study the dynamics of diffraction, in particular the transition
from soft to hard QCD. Several different scales are present in this
process, such as the $\gamma p$ centre-of-mass energy $W$, the photon
virtuality $Q^2$, the squared four-momentum transfer $t$ at the proton
vertex, and the mass $m_{V}$ of the vector meson, which can all be
tuned in the measurement (Figure \ref{fig:vm}). In a nonperturbative
description, the photon fluctuates into a vector meson that scatters
on the proton via soft Pomeron exchange, which predicts the energy
dependence of the $\gamma p$ cross section to be
$W^{4(\alphapom(t)-1)} \sim W^{0.22}$. This approach is able to
describe the data for light vector mesons if both $Q^2$ and $t$ are
small.  In the presence of a hard scale $Q^2$ or $m_V$, perturbative
QCD approaches based on two-gluon exchange predict a stronger rise of
the cross section with energy $W^2\sim 1/x$, corresponding to the
small-$x$ rise of the gluon density in the proton.

\begin{figure}
\centering
\epsfig{file=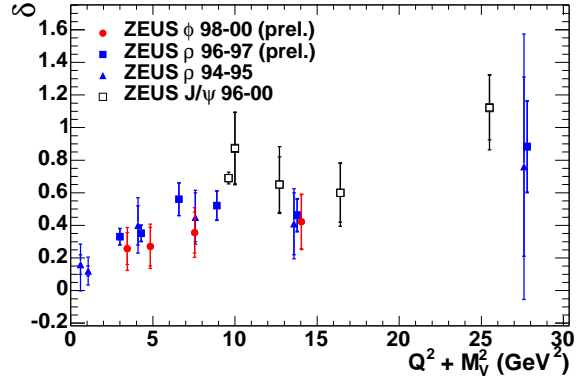,angle=270,clip=,width=0.59\linewidth}
\caption{\label{fig:zeusvma} $W^\delta$ dependence vs. $Q^2+m_{V}^2$
for elastic vector meson production from ZEUS.}
\end{figure}

\begin{figure}
\centering
\epsfig{file=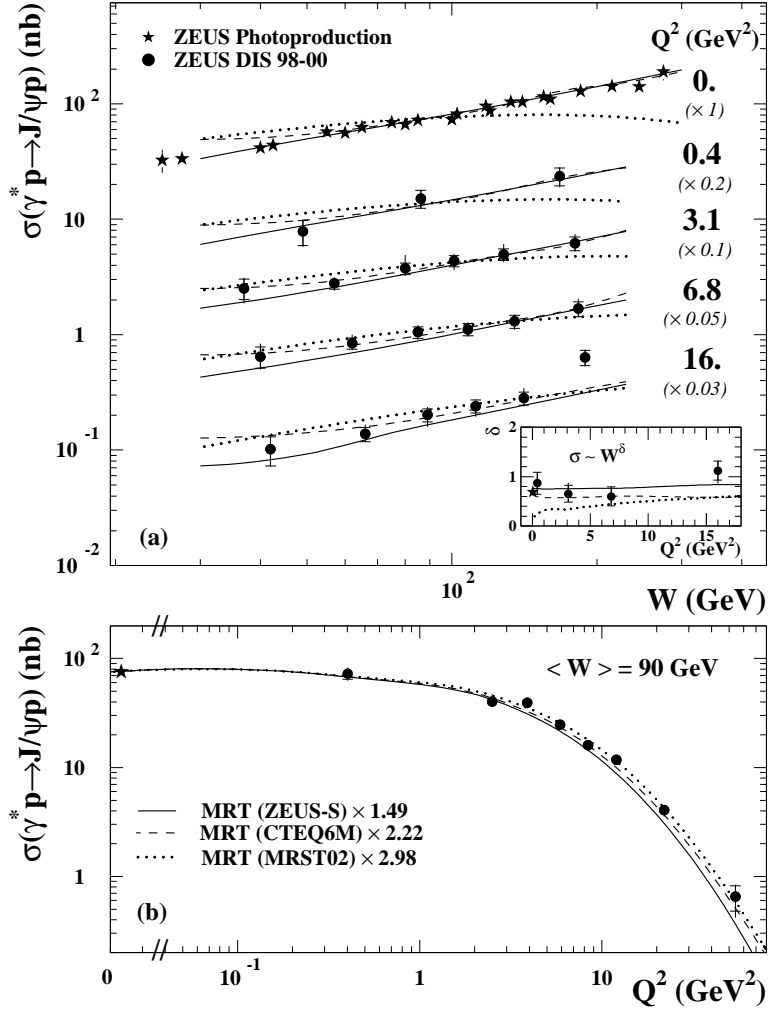,clip=,width=0.8\linewidth}
\caption{\label{fig:zeusvmb} $J/\Psi$ cross section dependence on $W$
and $Q^2$ from ZEUS.  The different curves correspond to different
gluon densities (see key in lower panel) used to model the generalized
gluon distribution.}
\end{figure}

ZEUS has presented \cite{helbich} new high statistics results on
exclusive $\phi$ production in DIS in the kinematic range $2<Q^2<70
\rm\ GeV^2$, $35<W<145 \rm\ GeV$ and $|t|<0.6 \rm\ GeV^2$.  The energy
dependence $W^\delta$ is observed to become steeper with $Q^2$:
$\delta$ increases from $0.26\pm0.10$ at $Q^2=2.4 \rm\ GeV^2$ to
$0.42\pm0.16$ at $Q^2=13 \rm\ GeV^2$. Thus, a picture emerges that at
around $Q^2 \sim m_{J/\Psi}^2\sim 10 \rm\ GeV^2$ the energy dependence
of the light vector mesons $\rho$ and $\phi$ becomes similar to the
one of the $J/\Psi$ at $Q^2\sim0$ (Figure \ref{fig:zeusvma}),
suggesting that $\mu^2 \sim Q^2+m^2_{V}$ could play the role of a
universal scale in this process.  Furthermore, from the $W$ dependence
measured in different $t$ intervals, the slope parameter of the
Pomeron trajectory $\alphapom'$ was extracted and found to be
consistent with zero, in contrast with the soft Pomeron model.  ZEUS
has also measured the $Q^2$ dependence of the exponential $t$ slope
parameter $b$, where some indication for a decrease with $Q^2$ is
observed.

Diffractive $J/\Psi$ production involves the charm quark mass as a
hard scale. This allows for a description in terms of hard-scattering
factorization and makes the process sensitive to the generalized gluon
distribution.  ZEUS \cite{bruni} has presented comprehensive results
on $J/\Psi$ production \cite{zeuspsi} for $0<Q^2<100 \rm\ GeV^2$ and
$30<W<220 \rm\ GeV$. In particular, cross sections as a function of
$W$ and $Q^2$ have been measured and compared with calculations based
on two-gluon exchange. The results exhibit a strong sensitivity to the
choice of gluon distribution in the calculation (Figure
\ref{fig:zeusvmb}). The Pomeron trajectory at $Q^2=7 \rm\ GeV^2$ was
extracted as $\alpha_\pom(t)= (1.20\pm0.03) + (0.07\pm0.05) t$, which
is substantially different from the soft Pomeron for both intercept
and slope.

Szymanowski \cite{szyman} has a presented the first NLO calculation
for exclusive $J/\Psi$ photoproduction \cite{ivanov} and for
electroproduction of light vector mesons in the framework of
leading-twist collinear factorization.  The dependence of the NLO
result on the factorization and renormalization scales is smaller than
at LO, as is generally expected.  The size of the NLO corrections in
$J/\Psi$ production is found to be substantial over a large kinematic
range.  The origin of this is currently not understood and further
study is needed.  One may speculate that the situation is better for
$J/\Psi$ electroproduction at sufficiently large $Q^2$, but an NLO
calculation of this process is presently not available.

$J/\Psi$ photoproduction with proton dissociation at high $|t|$
receives particular interest, because it offers a test of BFKL
dynamics in the exchanged gluon ladder.  Both H1 and ZEUS have
presented \cite{olsson} results on $J/\Psi$ production going up to
large values of $|t|$ \cite{h1hight,zeushight}.  H1 has measured the
cross section for $\gamma p \rightarrow J/\Psi\, Y$ up to $|t|=25 \rm\
GeV^2$ and found a power law behaviour in $t$ of $d\sigma/dt \sim
|t|^{-3.0\pm0.1}$. A BFKL calculation including an estimate of
next-to-leading corrections is in reasonable agreement with the H1 and
ZEUS data. However, the agreement becomes worse if effects of the
running of $\alpha_s$ are included.  Motyka \cite{motyka} has
presented a detailed theoretical study of vector meson photoproduction
at large $t$ in the BFKL framework, including the meson polarization
\cite{enberg}.

\begin{figure}
\centering
\vspace{1em}
\epsfig{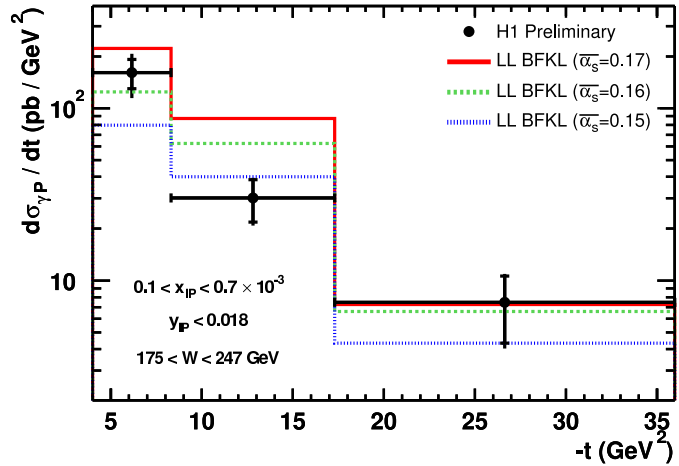}
\caption{\label{fig:photonsa} Diffractive high-$|t|$ photon
production cross section from H1.  }
\end{figure}

\begin{figure}
\centering
\epsfig{file=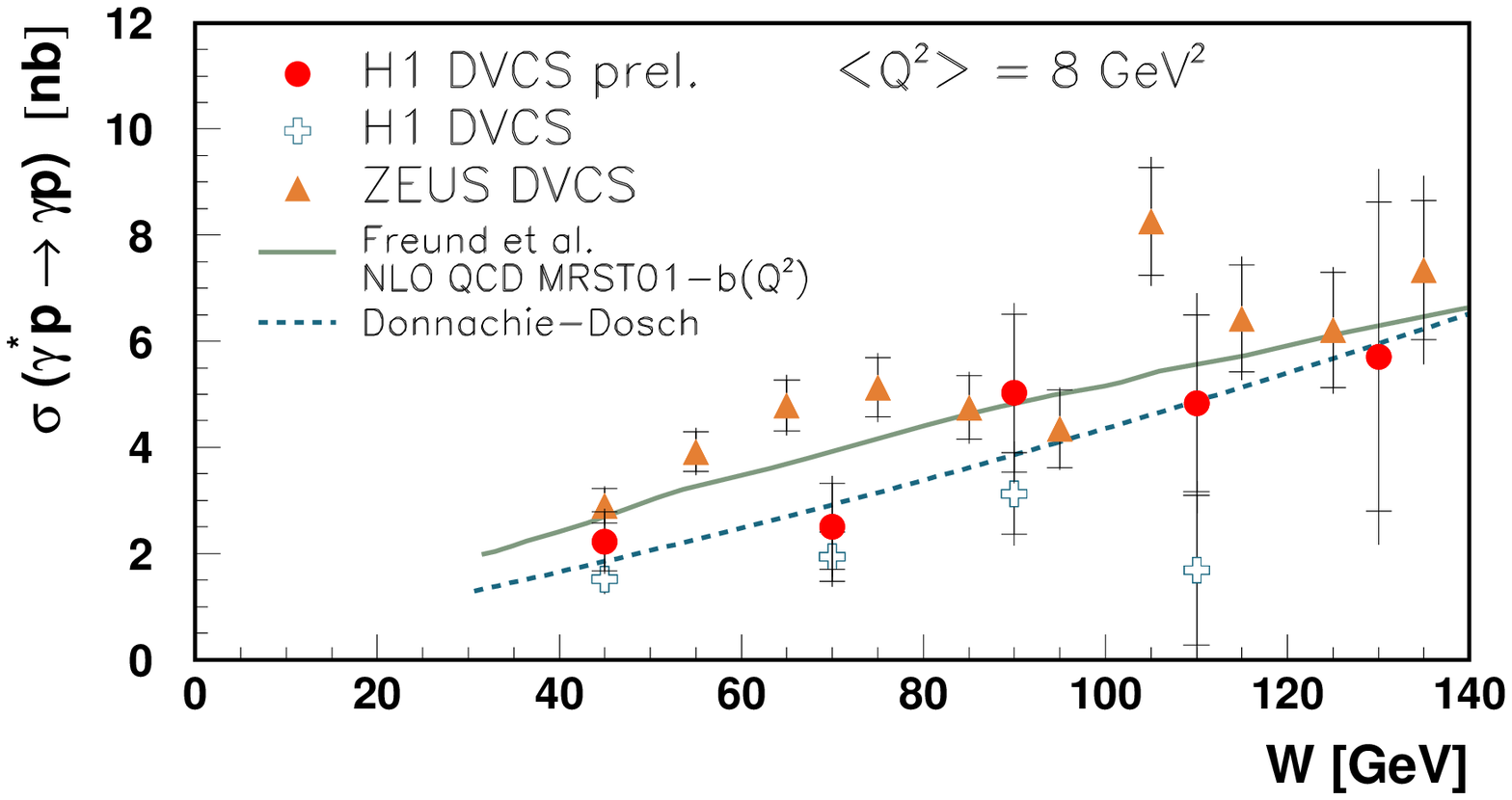,width=0.9\linewidth}
\epsfig{file=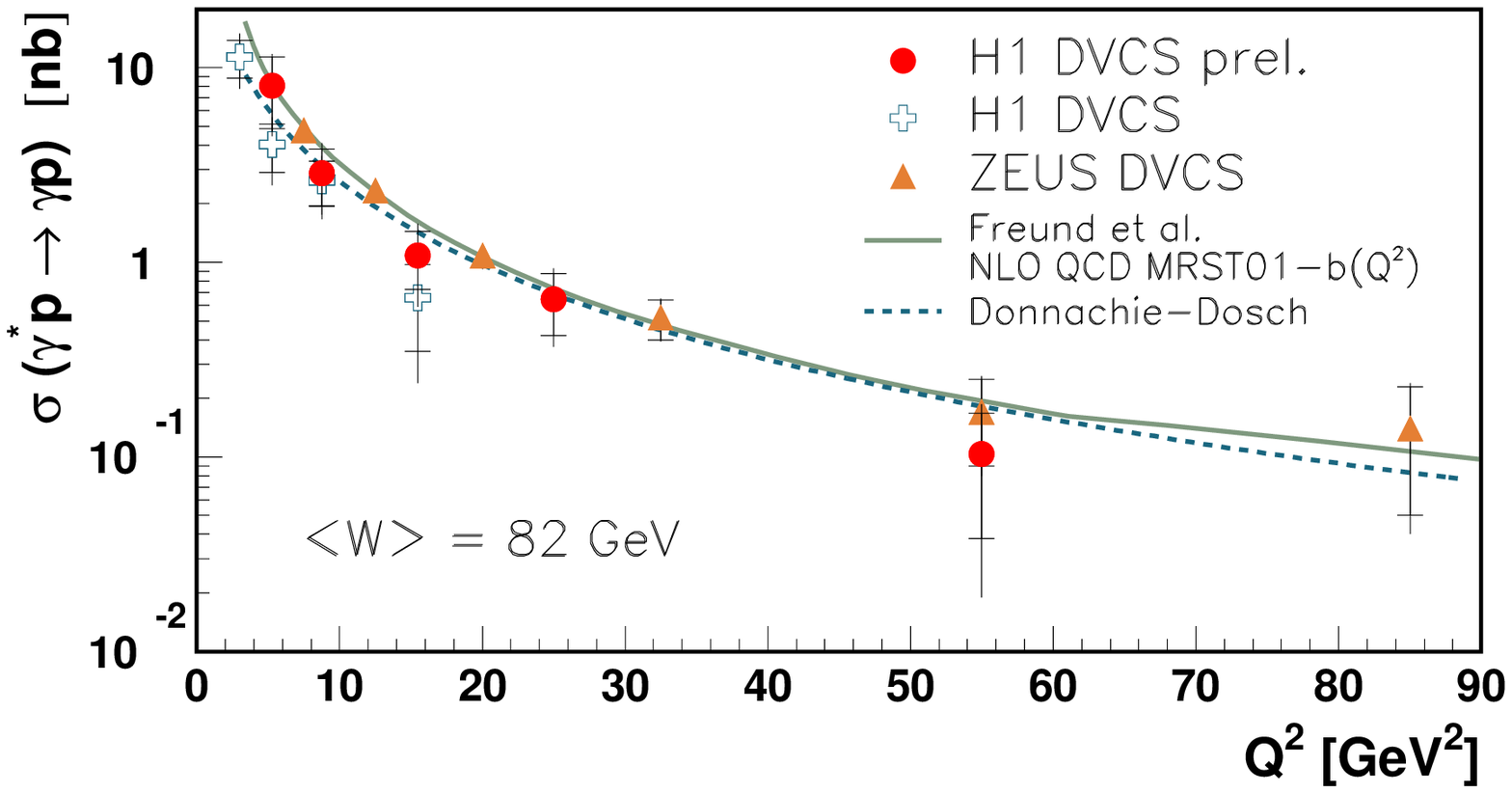,width=0.9\linewidth}
\caption{\label{fig:photonsb} DVCS cross section data compared with
  calculations in the framework of generalized parton distributions
  (Freund et al.) or of the dipole formulation (Donnachie and Dosch).  }
\end{figure}

An even simpler process to look for effects of BFKL evolution is
high-$|t|$ photon production $\gamma p \rightarrow \gamma Y$, which
does not involve a vector meson wave function.  H1 has presented
\cite{olsson} the first cross section measurement of this process in
photoproduction \cite{h1hightgamma}, reaching $|t|<35 \rm\ GeV^2$
(Figure \ref{fig:photonsa}).  The data are qualitatively described by
a leading-log BFKL calculation.

\begin{figure}
\centering
\epsfig{file=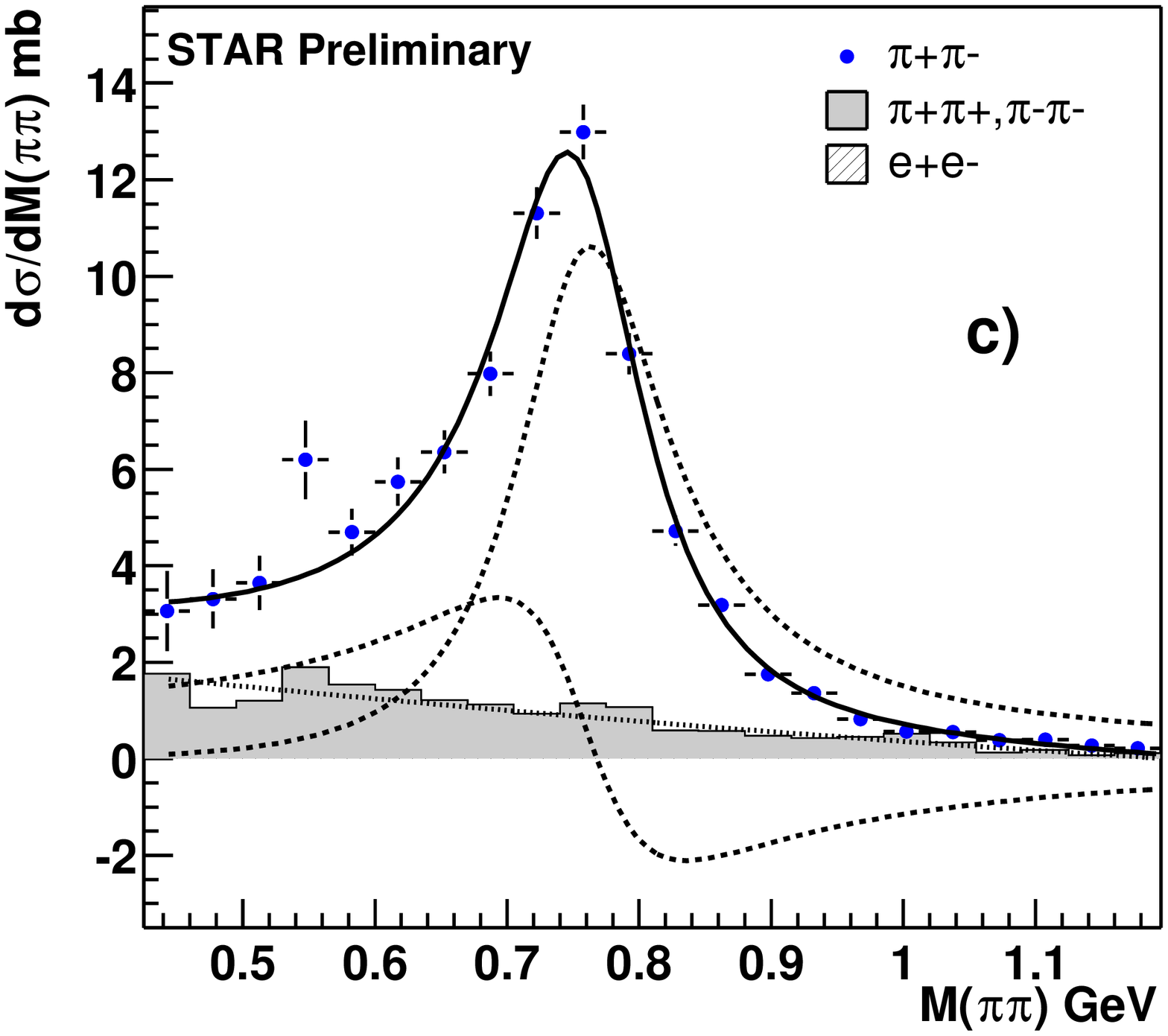,angle=0,clip=,width=0.65\linewidth}
\caption{\label{fig:star} $\rho^0$ peak in the $\pi\pi$ invariant mass
spectrum from Au+Au collisions at STAR.  }
\end{figure}

Deeply virtual Compton scattering (DVCS), i.e.\ exclusive photon
production $\gamma^* p \rightarrow \gamma p$ at small $|t|$ but large
$Q^2$, provides an opportunity to obtain information about generalized
parton distributions (GPDs).  Since the DVCS final state is
indistinguishable from the final state of the Bethe-Heitler process,
where the photon is radiated from the lepton, the amplitudes of the
two processes add coherently.  The resulting interference term is
estimated to be negligible in the kinematics of H1 and ZEUS, as long
as the measurement integrates over the azimuthal angle between the
lepton and the hadron plane \cite{bmk}. H1 and ZEUS have extracted
DVCS cross sections by subtracting the calculated Bethe-Heitler cross
section from the measured data \cite{dvcsh1,dvcszeus}.

H1 has presented \cite{favart} new results on the DVCS \cite{h1dvcs}
cross section for $4<Q^2<80\rm\ GeV^2$, $30<W<140 \rm\ GeV$ and
$|t|<1.0 \rm\ GeV^2$ (Figure \ref{fig:photonsb}).  The measured cross
sections as functions of $Q^2$ or $W$ are in reasonable agreement with
previous results from H1 and ZEUS.  They are compared with NLO QCD
calculations using GPD parameterizations based on the CTEQ6 and
MRST2001 PDFs, as well as with colour dipole models. All models
describe the data within the uncertainties if the $t$ slope parameter
is taken to be $b=7 \rm\ GeV^{-2}$. To constrain the models further,
it is necessary to measure the $t$-dependence of the DVCS cross
section.  This became also manifest in the presentation by Favart
\cite{favart-theo}, who compared the results of several saturation
models for DVCS: the present uncertainty in the cross section due to
the unknown $t$ dependence is in fact of similar size as the variation
of different models for saturation effects.

The $t$ dependence of DVCS and of exclusive vector meson production
has been a recurring theme at the workshop, with a dedicated
presentation by Weiss \cite{weiss}.  By a Fourier transform it can be
translated into information in the impact parameter plane.  In the
leading-twist formalism, the Fourier transform of generalized parton
distributions gives the transverse distance of the struck parton from
the proton center.  In the dipole formulation of the leading order
BFKL formalism, the Fourier transform of the scattering amplitude with
respect to $t$ gives the distance from the proton center where the
dipole scatters off gluons in the target.  This is of special
importance in connection with saturation (see section~\ref{sec:sat}).
Weiss \cite{weiss} has pointed out that partons with moderate momentum
fraction in the proton have a rather narrow impact parameter profile
and presented ideas how this might be used in studies of $pp$
collisions at the LHC \cite{frastrwei}.

The exclusive production of vector mesons can also be studied in
ultra-peripheral heavy ion collisions at RHIC. STAR presented
\cite{ogawa} results on diffractive $\rho^0$ production in Au+Au and
d+Au collisions, where a quasi-real photon emitted by one nucleus
scatters quasielastically off the other nucleus and becomes a vector
meson. STAR has studied samples where the interacting nuclei either
stay intact or dissociate because of nuclear excitation, which leads
to decay neutrons that are used to tag interactions at smaller impact
parameter.  The $\rho^0$ mass peak (Figure~\ref{fig:star}) observed by
STAR is similar in shape to $\rho^0$ photoproduction at HERA, and the
rapidity distribution matches a soft Pomeron model calculation.  The
$t$ dependence in Au+Au collisions is found consistent with an
interference effect due to the two indistinguishable production
diagrams.  The exponential $t$ slope parameter $b\approx 11.5 \rm\
GeV^{-2}$ obtained in d+Au collisions where the deuteron dissociates
is similar to values measured in scattering off the proton at HERA.

A theoretical study of the role of multiphoton exchange in lepton pair
production in heavy-ion collisions has been presented by Kuraev
\cite{kuraev}.


\section{Saturation}
\label{sec:sat}

A significant fraction of theoretical contributions in the working
group was devoted to saturation.  They were presented in a common
session with the working group on structure functions and low $x$,
which was complemented by a discussion session.  Studies oriented
towards phenomenology are reviewed in this section, whereas more
theoretical work is summarized in \cite{anna}.

A point of view expressed by several participants in the discussion
session is that the phenomenological success of colour dipole models
in describing HERA data does \emph{not} imply that the relevance of
saturation dynamics in HERA kinematics has been firmly established.
In this context it is important to distinguish between the breakdown
of the leading-twist (or ``DGLAP'') description, the onset of BFKL
dynamics, and the onset of saturation, by which we understand dynamics
involving nonlinear effects, strong gluon fields, and unitarisation of
the scattering amplitude.  In the colour dipole formulation,
saturation becomes relevant for dipoles of size $r \gsim 1/Q_s$, where
$Q_s$ is the saturation scale.  Typical dipole sizes $r \sim 1/Q$ are
selected by the momentum scale $Q$ of the process in question.  It was
felt by several participants that to establish saturation
convincingly, one needs processes where $Q$ is below $Q_s$ but at the
same time large enough to justify the use of perturbation theory in
the calculation.  Unfortunately, the strongest effects of saturation
are often found for $Q^2$ so small that the internal consistency of
perturbative arguments is not evident.  An example of such a situation
is the study of high-mass diffractive photon dissociation presented by
Munier \cite{munier}.  As an indicator of how sensitive a calculation
is to nonperturbative effects (and hence of the extent to which one
goes from theory to modelling) one may for instance take the
sensitivity of observables on the light quark mass which is often
included in the $q\bar{q}$ wave function of the photon.  This was done
in a study presented by Rogers \cite{rogers}, which in the framework
of a dipole model estimated that saturation effects are only of little
importance in the inclusive structure function $F_2$ at $x_B =
10^{-4}$ and $Q^2= 2~\mathrm{GeV}^2$, to give a concrete kinematical
point \cite{rogersetal}.

An important point is that the saturation scale $Q_s$ depends not only
on the energy variable $1/x$ but also on the impact parameter $b$ of
the scattering process.  Whereas inclusive observables such as the
structure functions $F_2$ and $F_L$ average over all impact
parameters, processes where the proton remains intact, like open
diffraction, vector meson production, or DVCS offer the possibility to
select the region of small impact parameters (corresponding to large
$t$).  In this region the gluon density in a proton is highest so that
saturation will set in earlier at a given $x$.  It is a renewed task
for theory to indicate which processes and kinematics are most
favorable in terms of sensitivity to saturation, of theoretical
control, and of experimental feasibility.

To include impact parameter dependence in the theoretical description
of saturation is challenging.  Investigations of this dependence are
at a rather early stage, and often it is still neglected altogether
(see \cite{anna}).  This also holds for the work reported by Utermann
\cite{utermann}, where the large gluon fields involved in saturation
are described by QCD instantons \cite{utermannschr}, thus providing a
new implementation of the idea that high-energy scattering can be
linked with fundamental properties of the QCD vacuum.


\section{Future Opportunities}
\label{sec:lhc}

Studies of diffractive phenomena represent an active field in
experimental particle physics at present as well as future hadronic
colliders. Both HERA and the TEVATRON are currently in their ``Run 2''
phases with upgraded detectors, for example the new H1 Very Forward
Proton Spectrometer VFPS \cite{janssen}, which provides full
acceptance for elastic protons in the diffractive regime around
$\xpom\sim 0.01$, or the new D0 Roman Pot system \cite{edwards}. A
wealth of precise data is still expected from these experiments.

In 2007, the Large Hadron Collider LHC is scheduled to start operation
at CERN, providing $pp$ collisions at $14 \rm\ TeV$ centre-of-mass
energy.  There are plans for diffractive studies using both
omni-purpose LHC experiments, ATLAS and CMS.  The TOTEM collaboration
\cite{deile} will install additional detectors around the CMS
interaction point, providing acceptance up to very large rapidities:
two tracking telescopes T1 and T2 for measurements of forward particle
production, which will cover $3.1<\eta<4.7$ and $5.3<\eta<6.5$, and a
set of Roman Pot spectrometers between $z=147\rm\ m$ and $z=220 \rm\
m$ on either side of the interaction point. The physics objectives
\cite{totemtdr} of TOTEM in standalone mode (without the CMS central
detector) are a measurement of the $pp$ elastic scattering cross
section $d\sigma/dt$ in the range $10^{-3}< |t| <10 \rm\ GeV^2$ and
the extraction of the total cross section at $14 \rm\ TeV$ with an
uncertainty of $1\%$ using the optical theorem.  These measurements
require only a few days of LHC running (not taking into account
commissioning) with a special high $\beta^*=1540 \rm\ m$ optics at low
luminosity $\mathcal{L}=10^{28} \rm\ cm^{-2}s^{-1}$.  In this
configuration, about $90\%$ of all diffractive protons are seen in the
Roman Pots, their momentum being measurable with a resolution around
$10^{-3}$.

For the nominal LHC optics ($\beta^*=0.5 \rm\ m$) and design
luminosity $\mathcal{L}>10^{33} \rm\ cm^{-2}s^{-1}$, there is a full
diffractive physics programme of TOTEM together with CMS
\cite{tasevsky}, with TOTEM being fully integrated into the CMS
trigger and data acquisition system as a CMS subdetector, in
particular providing an L1 trigger signal from the Roman Pots. The
programme includes single and double diffraction, hard diffraction
with jets, $W$s, or heavy quarks, search for diffractive SM or SUSY
Higgs production and other new physics, as well as studies of low-$x$
dynamics in the forward region \cite{cmseoi}.  The acceptance region
in $\xpom$ for the Roman Pots at $220 \rm\ m$ is $0.02 < \xpom < 0.2$.
The large centre-of-mass energy of the LHC provides an extension of
the accessible kinematic range to probe the structure of the
diffractive exchange to very low $\beta$ as well as to high $Q^2$.

In recent years, the possibility of discovering the Higgs boson in
double diffractive events at the LHC has gained significant interest,
because it provides in principle a clean process with a good signal
over background ratio, with the potential of a precise mass
reconstruction using the Roman Pots. However, to discover a light SM
or MSSM Higgs, additional Roman Pot detectors need to be installed at
$z=320$ and/or $z=420 \rm\ m$ to provide the necessary acceptance at
low $\xpom$.  Unfortunately, this would require installation in the
cold section of the LHC machine as well as more sophisticated trigger
scenarios since the Pots would be too far away from the interaction
region in order to provide an L1 trigger in time. The possible
installation of these additional Pots as well as other scenarios are
still under investigation.

A detailed analysis of diffractive Higgs production has been presented
by Royon \cite{royon}, where model predictions for signal and
background reactions were at the basis of simulations for the
CMS+TOTEM detector setup.  Critical issues for finding a standard
model Higgs identified in this study are triggering, cuts to enhance
the signal to background ratio, and the resolution in the missing mass
of the central system measured by the Roman pots.

The signal cross section used in the study of the Saclay group
\cite{royon} are in overall agreement with those presented by Martin
\cite{martin}, who gave an overview of the theory developed by the
Durham group to describe diffractive Higgs production and similar
processes in hadron-hadron collisions.  The generalized gluon
distribution (see section~\ref{sec:vmdvcs}) is an essential ingredient
in this approach.  As the Higgs would be detected in its $b\bar{b}$
decay mode, a major background to its observation is the diffractive
production of $b\bar{b}$ pairs via two-gluon fusion, which can be
described in the same theoretical framework.  The study of exclusive
dijet events, $p\bar{p}\to p + \mathrm{dijet} + \bar{p}$ at the
TEVATRON can provide valuable checks of theory approaches.  This is
especially important since the description of rapidity gap survival
has to rely on phenomenological models.

The ATLAS collaboration \cite{boonekamp} has very recently submitted a
Letter of Intent \cite{atlasloi} for the installation of additional
forward detectors, aimed at a precise luminosity determination.  Since
the ATLAS detector does not have sufficient forward coverage to
measure the total inelastic rate precisely enough, the collaboration
follows a different approach than TOTEM, namely using the very
challenging method of Coulomb scattering.  It is planned to install
Roman Pot detectors at $z=240 \rm\ m$ on either side of the ATLAS main
detector in order to measure elastic protons in the Coulomb region at
very small $|t|<5 \cdot 10^{-4} \rm\ GeV^2$ during a special high
$\beta^*$ optics LHC run at low luminosity.  This determination of
the absolute luminosity, which is planned to be precise within less
than $2\%$, would then be used to calibrate a luminosity monitor
(``Lucid'') based on Cerenkov counters placed around the beam pipe
close to the interaction point, which would provide luminosity
measurements for the standard LHC running.  In the future, this
programme will be extended to also cover diffractive and low-$x$
physics, similarly to the CMS+TOTEM plans.


\section{Conclusions}
\label{sec:end}

Diffraction has proven to be a healthy field at this workshop, with
theoretical activity on a wide range of issues, and data that brought
significant improvement over previous results or represented
altogether new measurements.  Preparation of studying diffractive
physics at the LHC is well under way.


\section*{Acknowledgements} 

We thank all participants in our session for the excellent
presentations and the lively discussions.  Special thanks are due to
Anna Stasto for co-organizing the common session on saturation and to
Larry McLerran for animating the discussion session on that topic.
The organisers of DIS 2004 deserve credit for an enjoyable and very
well organised meeting in pleasant surroundings.


\def\theseproc{{\it these proceedings}}


\end{document}